\documentclass[aps,superscriptaddress,prl,fleqn,showpacs,nofootinbib,preprintnumbers]{revtex4}

\usepackage{amssymb,amsmath,epsfig,slashed}

\begin{document}

\title{Exclusive double quarkonium production and generalized TMDs of gluons}

\author{Shohini Bhattacharya}
\affiliation{Department of Physics, SERC,
             Temple University, Philadelphia, PA 19122, USA}

\author{Andreas Metz}
\affiliation{Department of Physics, SERC,
             Temple University, Philadelphia, PA 19122, USA}

\author{Vikash Kumar Ojha}
\affiliation{School of Physics and Key Laboratory of Particle Physics and Particle Irradiation (MOE),
              Shandong University, Jinan, Shandong 250100, China}              

\author{Jeng-Yuan Tsai}
\affiliation{Department of Physics, SERC,
             Temple University, Philadelphia, PA 19122, USA}             
                           
\author{Jian Zhou}
\affiliation{School of Physics and Key Laboratory of Particle Physics and Particle Irradiation (MOE),
              Shandong University, Jinan, Shandong 250100, China}
              
\begin{abstract}
Being the ``mother distributions" of all types of two-parton correlation functions, generalized transverse momentum dependent parton distributions (GTMDs) have attracted a lot of attention over the last years. 
We argue that exclusive double production of pseudoscalar quarkonia ($\eta_c$ or $\eta_b$) in nucleon-nucleon collisions gives access to GTMDs of gluons.
\end{abstract}


\date{\today}

\maketitle

\section{Introduction} 
\label{s:intro}
Generalized transverse momentum dependent parton distributions (GTMDs) for hadrons are often denoted as ``mother distributions" as they represent the most general (two-)parton correlation functions of hadrons~\cite{Ji:2003ak, Belitsky:2003nz, Meissner:2008ay, Meissner:2009ww, Lorce:2013pza}.
They allow one to study the multi-dimensional parton structure of hadrons at the next level. 
In this context we recall that all generalized parton distributions (GPDs) and transverse momentum dependent parton distributions (TMDs) are kinematical projections of certain GTMDs.

The ``mother distribution" character of GTMDs provides a strong motivation for exploring these quantities.
But several further motivations exist.
First, for a vanishing longitudinal momentum transfer to the hadron, the Fourier transform of a GTMD is a Wigner function that depends on the (average) longitudinal and transverse momentum as well as transverse position of partons inside the hadron~\cite{Belitsky:2003nz, Lorce:2011dv}.
Wigner functions, which are also popular in other areas like AMO physics, are the quantum mechanical counterpart of classical phase space distributions~\cite{Wigner:1932eb}.
Partonic Wigner functions in principle allow one to make 5-D images of hadrons --- see for instance Refs.~\cite{Lorce:2011dv, Lorce:2011kd, Lorce:2015sqe}.
Second, for both quarks and gluons there is a very intriguing relation between a specific GTMD --- $F_{1,4}$ in the notation of~\cite{Meissner:2009ww} --- and the orbital angular momentum (OAM) of partons inside a longitudinally polarized nucleon~\cite{Lorce:2011kd, Hatta:2011ku, Kanazawa:2014nha, Hagler:2003jw}.
In the quark case this relation applies for the two commonly used definitions of OAM~\cite{Hatta:2011ku, Ji:2012sj, Lorce:2012ce}  --- the canonical OAM by Jaffe and Manohar ($L_{\rm JM}$)~\cite{Jaffe:1989jz} and the kinetic OAM by Ji ($L_{\rm Ji}$)~\cite{Ji:1996ek} --- and it allows for an intuitive interpretation of the difference $L_{\rm JM} - L_{\rm Ji}$~\cite{Burkardt:2012sd}.
It also gives access to the so far elusive $L_{\rm JM}$ in quantum chromodynamics (QCD) on the lattice~\cite{Hatta:2011ku, Rajan:2016tlg, Engelhardt:2017miy}.
Third, certain GTMDs are related to spin-orbit correlations of the nucleon, where the best studied cases are for longitudinal polarization of either the parton or the target~\cite{Lorce:2011kd, Lorce:2014mxa, Kanazawa:2014nha}. 
Such spin-orbit correlations have a meaning similar to the ones in an atomic system such as the hydrogen atom.

A number of model calculations of GTMDs is available by
now~\cite{Meissner:2008ay, Meissner:2009ww, Lorce:2011dv, Lorce:2011kd, Lorce:2011ni, Kanazawa:2014nha, Mukherjee:2014nya, Hagiwara:2014iya, Lorce:2015sqe, Chakrabarti:2016yuw, Echevarria:2016mrc, Hagiwara:2016kam, Gutsche:2016gcd, Zhou:2016rnt, Courtoy:2016des, More:2017zqq, More:2017zqp}.
On the other hand, for more than a decade after the first papers on GTMDs~\cite{Ji:2003ak, Belitsky:2003nz} had appeared it was not known how these functions could be measured.
Only recently has it been argued that gluon GTMDs can be accessed through exclusive hard diffractive di-jet production in DIS~\cite{Hatta:2016dxp}, a reaction which should be measurable at a future electron-ion collider~\cite{Boer:2011fh, Accardi:2012qut}.
That work on GTMD observables was followed by several related studies~\cite{Zhou:2016rnt, Ji:2016jgn, Hatta:2016aoc, Hagiwara:2017ofm, Iancu:2017fzn, Bhattacharya:2017bvs, Hagiwara:2017fye}, which all but one~\cite{Bhattacharya:2017bvs} deal with GTMDs of gluons.

With the exception of Ref.~\cite{Ji:2016jgn}, where a very specific weighted cross section was explored, all the works on observables for gluon GTMDs focus on the small-$x$ region of parton saturation. 
Here we show that GTMDs of gluons, for any momentum fraction $x$, can be addressed through exclusive double production of pseudoscalar quarkonia ($\eta_c$ or $\eta_b$) in nucleon-nucleon collisions, $N_a N_b \to \eta_Q \eta_Q N_a N_b$, where $\eta_Q$ denotes either $\eta_c$ or $\eta_b$. 
We discuss how specific GTMDs can be accessed by means of suitable polarization observables.
Our analysis has similarities to the study in Ref.~\cite{Bhattacharya:2017bvs} where we have shown that quark GTMDs can in principle be measured through the exclusive pion-nucleon double Drell-Yan process, $\pi N \to (\ell_1^- \ell_1^+) (\ell_2^- \ell_2^+) N'$.

\section{Generalized TMDs of gluons} 
\label{s:gtmd}
A complete list of gluon GTMDs for a spin-$\frac{1}{2}$ target was presented in Ref.~\cite{Lorce:2013pza}.
They can be defined through the off-forward gluon-gluon correlator
\begin{equation} 
W_{\lambda,\lambda'}^{g \, [i j]} (P,\Delta,x,\vec{k}_\perp) = 
\frac{1}{P^+} \int \frac{dz^- \, d^2\vec{z}_\perp}{(2\pi)^3} \, e^{i k \cdot z} \, 
\langle p', \lambda' | \, F_a^{+ i}(- \tfrac{z}{2}) \, {\cal W}_{ab} (- \tfrac{z}{2}, \tfrac{z}{2}) \,  F_b^{+ j}(\tfrac{z}{2})\, | p, \lambda \rangle \Big|_{z^+ = 0} \,,
\label{e:gtmd_corr}
\end{equation}
which is a generalization of the correlator for gluon GPDs to include transverse parton momenta --- see for instance Ref.~\cite{Meissner:2007rx}.
In Eq.~\eqref{e:gtmd_corr}, the gluons are represented by components of the field strength tensor $F_a^{\mu \nu}$, where `$a$' is a color index.
Note that we have limited ourselves to the case of leading twist --- $i, j$ are transverse indices, and `$+$' indicates the light-cone plus component.
(The light-cone components of a generic 4-vector $a = (a^0, a^1, a^2, a^3)$ are defined through $a^\pm = (a^0  \pm a^3) / \sqrt{2}$ and $\vec{a}_\perp = (a^1, a^2)$.) 
The Wilson line ${\cal W}_{ab}$, which is in the adjoint representation, ensures color gauge invariance of the bi-local gluon operator.
The initial (final) nucleon is specified through its four-momentum $p \; (p')$ and helicity $\lambda \; (\lambda')$.
We also have used the average four-momentum $P = (p + p')/2$ and the momentum transfer $\Delta = p' - p$.
The average longitudinal and transverse gluon momenta are denoted by $x$ and $\vec{k}_\perp$, respectively.

At leading twist, one has a total of 16 gluon GTMDs~\cite{Lorce:2013pza}.
Here we concentrate on
\begin{eqnarray} 
W_{\lambda, \lambda'}^g & = & \delta_\perp^{i j} \, W_{\lambda,\lambda'}^{g \, [i j]}
\nonumber \\
& = & \frac{1}{2M} \, \bar{u}(p',\lambda') \bigg[ 
F_{1,1}^g + \frac{i  \sigma^{i+}  k_\perp^i}{P^+} \, F_{1,2}^g + \frac{i \, \sigma^{i+} \Delta_\perp^i}{P^+} \, F_{1,3}^g 
+ \frac{i \sigma^{ij} k_{\perp}^i \Delta_{\perp}^j}{M^2} \, F_{1,4}^g  \bigg] u(p,\lambda)
\nonumber \\
& = & \frac{1}{M \sqrt{1 - \xi^2}} \bigg\{ 
\bigg[ M \delta_{\lambda,\lambda'} - \frac{1}{2} \Big( \lambda \Delta_\perp^1 + i \Delta_\perp^2 \Big) \delta_{\lambda,-\lambda'} \bigg] F_{1,1}^g
+ (1 - \xi^2) \Big( \lambda k_\perp^1 + i k_\perp^2 \Big) \delta_{\lambda,-\lambda'} \, F_{1,2}^g
\nonumber \\
& & \hspace{0.8cm} 
+ \; (1 - \xi^2) \Big( \lambda \Delta_\perp^1 + i \Delta_\perp^2 \Big) \delta_{\lambda,-\lambda'} \, F_{1,3}^g
+ \frac{i \varepsilon_\perp^{ij} k_{\perp}^i \Delta_{\perp}^j}{M^2} \bigg[ \lambda M \delta_{\lambda,\lambda'} - \frac{\xi}{2} \Big( \Delta_\perp^1 + i \lambda \Delta_\perp^2 \Big) \delta_{\lambda,-\lambda'} \bigg] F_{1,4}^g \bigg\} \,,
\label{e:GTMD_unpol}
\\[0.2cm]  
\widetilde{W}_{\lambda, \lambda'}^g & = & - \, i \varepsilon_\perp^{i j} \, W_{\lambda,\lambda'}^{g \, [i j]}
\nonumber \\
& = & \frac{1}{2M} \, \bar{u}(p',\lambda') \bigg[ 
- \frac{i \varepsilon_\perp^{ij} k_{\perp}^i \Delta_{\perp}^j}{M^2} \, G_{1,1}^g
+ \frac{i  \sigma^{i+}  \gamma_5 k_\perp^i}{P^+} \, G_{1,2}^g + \frac{i  \sigma^{i+}  \gamma_5 \Delta_\perp^i}{P^+} \, G_{1,3}^g
+ i \sigma^{+-} \gamma_5 \, G_{1,4}^g  \bigg] u(p,\lambda)
\nonumber \\
& = & \frac{1}{M \sqrt{1 - \xi^2}} \bigg\{
- \frac{i \varepsilon_\perp^{ij} k_{\perp}^i \Delta_{\perp}^j}{M^2} \bigg[ M \delta_{\lambda,\lambda'} - \frac{1}{2} \Big( \lambda \Delta_\perp^1 + i \Delta_\perp^2 \Big) \delta_{\lambda,-\lambda'} \bigg] G_{1,1}^g
+ (1 - \xi^2) \Big( k_\perp^1 + i \lambda k_\perp^2 \Big) \delta_{\lambda,-\lambda'} \, G_{1,2}^g
\nonumber \\
& & \hspace{0.8cm} 
+ \; (1 - \xi^2) \Big( \Delta_\perp^1 + i \lambda \Delta_\perp^2 \Big) \delta_{\lambda,-\lambda'} \, G_{1,3}^g
+ \bigg[ \lambda M \delta_{\lambda,\lambda'} - \frac{\xi}{2} \Big( \Delta_\perp^1 + i \lambda \Delta_\perp^2 \Big) \delta_{\lambda,-\lambda'} \bigg] G_{1,4}^g \bigg\} \,.
\label{e:GTMD_hel}
\end{eqnarray}
The expression in~\eqref{e:GTMD_unpol} describes unpolarized gluons, while~\eqref{e:GTMD_hel} describes the gluon helicity distribution.
Below we will refer to the GTMDs defined in Eq.~\eqref{e:GTMD_unpol} and in Eq.~\eqref{e:GTMD_hel} as F-type and G-type GTMDs, respectively.
(For linearly polarized gluons one needs another eight GTMDs~\cite{Lorce:2013pza}.)
In Eqs.~\eqref{e:GTMD_unpol},~\eqref{e:GTMD_hel} we have used $\delta_\perp^{ij} = - g_\perp^{ij}$, where $g^{\mu \nu}$ denotes the metric tensor, and $\varepsilon_\perp^{ij} = \varepsilon^{-+ij}$ with $\varepsilon^{0123} = 1$.
The GTMD correlator $W_{\lambda, \lambda'}^g$ in~\eqref{e:GTMD_unpol} can be parameterized like the one for unpolarized quarks, while $\widetilde{W}_{\lambda, \lambda'}^g$ in~\eqref{e:GTMD_hel} can be parameterized like the one for longitudinally polarized quarks.
(For a corresponding discussion in the case of quark and gluons GPDs and TMDs we refer to~\cite{Meissner:2007rx}.)
Our notation for the gluon GTMDs therefore follows the quark case presented in~\cite{Meissner:2009ww}.
In order to evaluate the matrix elements in Eqs.~\eqref{e:GTMD_unpol},~\eqref{e:GTMD_hel} we considered $u(p,\lambda)$ and $u(p',\lambda')$ as light-cone helicity spinors \cite{Soper:1972xc, Diehl:2003ny}.
The nucleon mass is denoted by $M$, while $\xi = (p^+ - p'^+) / (p^+ + p'^+) = - \Delta^+ / (2 P^+)$ is the longitudinal momentum transfer to the nucleon.
Like TMDs, a generic GTMD $X = X(x, \xi, \vec{k}_\perp, \vec{\Delta}_\perp)$ depends on $x$ and $\vec{k}_\perp$.
Also, like GPDs, it depends on $\xi$ and $\vec{\Delta}_\perp$.
In general, GTMDs are complex-valued functions~\cite{Meissner:2008ay, Meissner:2009ww}.
For brevity we have not indicated the scale dependence of GTMDs --- see for instance Ref~\cite{Echevarria:2016mrc} for more details on this point.

When discussing observables below we will focus on the two gluon GTMDs $F_{1,4}$ and $G_{1,1}$. 
As already mentioned in the Introduction, $F_{1,4}$ gives access to the OAM of gluons, which may play an important role for the spin sum rule of the nucleon~\cite{Hatta:2016aoc, Hatta:2018itc}.
Both $F_{1,4}$ and $G_{1,1}$ quantify the strength of spin-orbit correlations~\cite{Lorce:2011kd, Lorce:2014mxa, Kanazawa:2014nha} --- the correlation between the spin of the nucleon and the OAM of the gluon in the case of $F_{1,4}$, and the correlation between the spin of the gluon and its OAM in the case of $G_{1,1}$.
Because of these reasons, $F_{1,4}$ and $G_{1,1}$ have attracted the most attention so far.

\section{Scattering amplitude} 
\label{s:amplitude}
Now we discuss the scattering amplitude of the exclusive process
\begin{equation}
N_{a}(p_a,\lambda_a) + N_{b}(p_b,\lambda_b) \to \eta_Q(q_1) + \eta_Q(q_2) + N_{a}(p_a',\lambda_a') + N_{b}(p_b',\lambda_b') \,,
\label{e:process}
\end{equation}
with $q_1^2 = q_2^2 = m_{\eta}^2$.
We assume $p_a^+$ and $p_b^-$ to be large, and we consider large $s = (p_a + p_b)^2 \approx 2 p_a^+ p_b^-$ as well as small transverse momenta of the quarkonia, $| \vec{q}_{i \perp} | \ll m_{\eta}$.
One can use TMD factorization in this kinematical region. 
The mass $m_{\eta}$ serves as the large scale that justifies a treatment in perturbative QCD.

A total of eight Feynman diagrams contribute to lowest order in the strong coupling constant $\alpha_s = g_s^2 / (4 \pi)$.
(Note that we focus on kinematics in which the emission of two $\eta_Q$ from a single gluon line connecting the two nucleons is suppressed.)
We find the following result for the scattering amplitude:
\begin{eqnarray} 
{\cal T}_{\lambda_a, \lambda_a' ; \lambda_b, \lambda_b'} & = & 
- \, i \, A
\int d^2 \vec{k}_{a \perp} \int d^2 \vec{k}_{b \perp} \, \delta^{(2)} \bigg( \frac{\Delta \vec{q}_{\perp}}{2} - \vec{k}_{a \perp} -\vec{k}_{b \perp} \bigg)
\nonumber \\[0.1cm]
&& \times \, \Big[ W_{\lambda_a, \lambda_a'}^g (x_a, \vec{k}_{a \perp}) \, W_{\lambda_b, \lambda_b'}^g (x_b, \vec{k}_{b \perp})
+ W_{\lambda_a, \lambda_a'}^g (- x_a, - \vec{k}_{a \perp}) \, W_{\lambda_b, \lambda_b'}^g (- x_b, - \vec{k}_{b \perp})
\nonumber \\[0.1cm]
& & + \, \widetilde{W}_{\lambda_a, \lambda_a'}^g (x_a, \vec{k}_{a \perp}) \, \widetilde{W}_{\lambda_b, \lambda_b'}^g (x_b, \vec{k}_{b \perp})
 + \widetilde{W}_{\lambda_a, \lambda_a'}^g (- x_a, - \vec{k}_{a \perp}) \, \widetilde{W}_{\lambda_b, \lambda_b'}^g (- x_b, - \vec{k}_{b \perp}) \Big]
\nonumber \\[0.1cm]
& = & - \, 2 i \, A 
\int d^2 \vec{k}_{a \perp} \int d^2 \vec{k}_{b \perp} \, \delta^{(2)} \bigg( \frac{\Delta \vec{q}_{\perp}}{2} - \vec{k}_{a \perp} -\vec{k}_{b \perp} \bigg)
\nonumber \\[0.1cm]
&& \times \, \Big[ W_{\lambda_a, \lambda_a'}^g (x_a, \vec{k}_{a \perp}) \, W_{\lambda_b, \lambda_b'}^g (x_b, \vec{k}_{b \perp})
+ \widetilde{W}_{\lambda_a, \lambda_a'}^g (x_a, \vec{k}_{a \perp}) \, \widetilde{W}_{\lambda_b, \lambda_b'}^g (x_b, \vec{k}_{b \perp}) \Big] \,,
\label{e:amplitude}
\end{eqnarray}
with the constant $A$ given by
\begin{equation}
A = \frac{g_s^4 \, R^{2}_{0}(0) \, s}{N_c (N_c^2 - 1) \, \pi \, m_{\eta}^5 \, (1 + \xi_a) (1+\xi_b)} \,.
\label{e:constant}
\end{equation}
In Eq.~\eqref{e:constant}, $R_0(0)$ is the value of the radial wave function of the $\eta_Q$ at the origin, and $N_c$ is the number of quark colors.
The longitudinal momentum transfer to the nucleons  is given by $\xi_a = (q_1^+ + q_2^+) / (2 P_a^+)$ and $\xi_b = (q_1^- + q_2^-) / (2 P_b^-)$.
In this leading-order calculation, the longitudinal momenta of the gluons are fixed according to $x_a = (q_1^+ - q_2^+) / (2 P_a^+)$ and $x_b = (q_1^- - q_2^-) / (2 P_b^-)$.
This means that one has access to the GTMDs in the ERBL region~\cite{Efremov:1979qk, Lepage:1979zb}, which is characterized by $- \xi \le x \le \xi$.
The transverse gluon momenta are integrated over with the constraint given by the delta function in Eq.~\eqref{e:amplitude}.
Note that we have used the transverse momentum $\Delta \vec{q}_T = \vec{q}_{1 \perp} - \vec{q}_{2 \perp}$.
Moreover, in order to obtain the last member in Eq.~\eqref{e:amplitude} we have exploited the symmetry properties 
$W_{\lambda, \lambda'}^g (x, \vec{k}_{\perp}) = W_{\lambda, \lambda'}^g (- x, - \vec{k}_{\perp})$,
$\widetilde{W}_{\lambda, \lambda'}^g (x, \vec{k}_{\perp}) = - \, \widetilde{W}_{\lambda, \lambda'}^g (- x, - \vec{k}_{\perp})$, which hold for both nucleons.
According to Eq.~\eqref{e:amplitude}, F-type and G-type GTMDs as well as their interference enter in the observables.

\section{Cross section and polarization observables} 
\label{s:observables}
The cross section in the center-of-mass frame is obtained from the scattering amplitude in Eq.~\eqref{e:amplitude} via
\begin{equation}
d \sigma_{\lambda_a, \lambda_a'; \lambda_b, \lambda_b'} 
= \frac{\pi}{2 s^{3/2}} \frac{1 + \xi_a}{1 - \xi_a} \, |{\cal T}_{\lambda_a, \lambda_a' ; \lambda_b, \lambda_b'}|^2 
\, \delta(p_a^{\prime 0} + p_b^{\prime 0} + q_1^0 + q_2^0 - \sqrt{s}) 
\, \frac{d^3 \vec{q}_1}{(2 \pi)^3 2 q_1^0} \, \frac{d^3 \vec{q}_2}{(2 \pi)^3 2 q_2^0} \, \frac{d^3 \vec{p}_b^{\, \prime}}{(2 \pi)^3 2 p_b^{\prime 0}} \,,
\label{e:cross_section}
\end{equation}
where we have already integrated over the phase space of one of the outgoing nucleons.
Note that Eq.~\eqref{e:cross_section} contains the symmetry factor $\frac{1}{2}$.
In defining polarization observables we focus on the GTMDs of the nucleon $N_a$, while we always sum/average over the spins of the nucleon $N_b$ by using $\frac{1}{2} \sum_{\lambda_b, \lambda_b'}$.
In the following we consider the unpolarized cross section, single-spin asymmetries (SSAs), and double-spin asymmetries (DSAs).
To this end we introduce
\begin{eqnarray} 
\tau_{UU} & = & \frac{1}{2} \sum_{\lambda_b, \lambda_b'}
\frac{1}{2} \sum_{\lambda_a, \lambda_a'}  |{\cal T}_{\lambda_a, \lambda_a' ; \lambda_b, \lambda_b'}|^2 \,,
\label{e:po_unpol}
\\[0.1cm]
\tau_{LU} & = & \frac{1}{2} \sum_{\lambda_b, \lambda_b'}
\frac{1}{2} \sum_{\lambda_a'} \Big( |{\cal T}_{+, \lambda_a' ; \lambda_b, \lambda_b'}|^2 - |{\cal T}_{-, \lambda_a' ; \lambda_b, \lambda_b'}|^2 \Big) \,,
\label{e:po_ssa}
\\[0.1cm]
\tau_{LL} & = & \frac{1}{2}\sum_{\lambda_b, \lambda_b'}
\frac{1}{2} \Big( \big( |{\cal T}_{+, + ; \lambda_b, \lambda_b'}|^2 - |{\cal T}_{+, - ; \lambda_b, \lambda_b'}|^2 \big)
- \big( |{\cal T}_{-, + ; \lambda_b, \lambda_b'}|^2 - |{\cal T}_{-, - ; \lambda_b, \lambda_b'}|^2 \big) \Big) \,,
\label{e:po_dsa}
\end{eqnarray}
where the indices in $\tau_{UU}$, $\tau_{LU}$ and $\tau_{LL}$ refer to the nucleon $N_a$.
The quantity in~\eqref{e:po_ssa} describes the numerator of the longitudinal target SSA, while the quantity in~\eqref{e:po_dsa} determines the longitudinal DSA. 
Analogous definitions apply for spin asymmetries involving transverse polarization in the $x$-direction or $y$-direction.

Below we will set $\vec{\Delta}_{b \perp} = 0$ throughout, which simplifies the expressions for the observables.
In this case one has two independent external transverse momenta only, for which we choose $\Delta \vec{q}_\perp$ and $\vec{\Delta}_{a \perp} = - (\vec{q}_{1 \perp} + \vec{q}_{2 \perp})$.

As discussed above, we are specifically interested in gaining access to $F_{1,4}$ and $G_{1,1}$. 
In order to address $F_{1,4}$, one can consider the following linear combination of polarization observables (see also Ref.~\cite{Bhattacharya:2017bvs}):
\begin{eqnarray}
\lefteqn{\frac{1}{4} \big( \tau_{UU} + \tau_{LL} - \tau_{XX} - \tau_{YY} \big)}
\nonumber \\[0.1cm]
& = & \frac{\varepsilon_\perp^{ij} \Delta_{a \perp}^j}{M} \, \frac{\varepsilon_\perp^{kl} \Delta_{a \perp}^l}{M} \,
C \bigg[ \frac{k_{a \perp}^i}{M} F_{1,4}(x_a, \vec{k}_{a\perp}) \, F_{1,1}(x_b, \vec{k}_{b\perp}) \bigg] \,
C \bigg[ \frac{p_{a \perp}^k}{M} F_{1,4}^{\ast}(x_a, \vec{p}_{a\perp}) \, F_{1,1}^{\ast}(x_b, \vec{p}_{b\perp}) \bigg]
\nonumber \\[0.1cm]
& + & (1 - \xi_b^2)^2 \, \frac{\varepsilon_\perp^{ij} \Delta_{a \perp}^j}{M} \, \frac{\varepsilon_\perp^{kl} \Delta_{a \perp}^l}{M} \,
C \bigg[ \frac{k_{a \perp}^i k_{b \perp}^m}{M^2} F_{1,4}(x_a, \vec{k}_{a\perp}) \, F_{1,2}(x_b, \vec{k}_{b\perp}) \bigg] \,
C \bigg[ \frac{p_{a \perp}^k p_{b \perp}^m}{M^2} F_{1,4}^{\ast}(x_a, \vec{p}_{a\perp}) \, F_{1,2}^{\ast}(x_b, \vec{p}_{b\perp}) \bigg]
\nonumber \\[0.1cm]
& + & C \Big[ G_{1,4}(x_a, \vec{k}_{a\perp}) \, G_{1,4}(x_b, \vec{k}_{b\perp}) \Big] \,
C \Big[ G_{1,4}^{\ast}(x_a, \vec{p}_{a\perp}) \, G_{1,4}^{\ast}(x_b, \vec{p}_{b\perp}) \Big]
\nonumber \\[0.1cm]
& + & (1 - \xi_b^2)^2 \, C \bigg[ \frac{k_{b \perp}^m}{M} G_{1,4}(x_a, \vec{k}_{a\perp}) \, G_{1,2}(x_b, \vec{k}_{b\perp}) \bigg] \,
C \bigg[ \frac{p_{b \perp}^m}{M} G_{1,4}^{\ast}(x_a, \vec{p}_{a\perp}) \, G_{1,2}^{\ast}(x_b, \vec{p}_{b\perp}) \bigg]
\nonumber \\[0.1cm]
& + & 2 (1 - \xi_b^2)^2 \, \frac{\varepsilon_\perp^{ij} \Delta_{a \perp}^j}{M} \, \varepsilon_{\perp}^{mn} \, {\rm Re} \,
\bigg\{ C \bigg[ \frac{k_{a \perp}^i k_{b \perp}^m}{M^2} F_{1,4}(x_a, \vec{k}_{a\perp}) \, F_{1,2}(x_b, \vec{k}_{b\perp}) \bigg] \,
C \bigg[ \frac{p_{b \perp}^n}{M} G_{1,4}^{\ast}(x_a, \vec{p}_{a\perp}) \, G_{1,2}^{\ast}(x_b, \vec{p}_{b\perp}) \bigg] \bigg\} \,, \qquad
\label{e:obs_F14}
\end{eqnarray}
where we have used the shorthand notation
\begin{eqnarray}
C \Big[ w(\vec{k}_{a\perp},\vec{k}_{b\perp}) \, X \, Y \Big] & = & 
\frac{2 A}{\sqrt{1 - \xi_a^2} \, \sqrt{1 - \xi_b^2}} \int d^2 \vec{k}_{a\perp} \int d^2 \vec{k}_{b\perp} \,
\delta^{(2)}\bigg( \frac{\Delta \vec{q}_\perp}{2} - \vec{k}_{a\perp} - \vec{k}_{b\perp} \bigg)  
\nonumber \\[0.2cm]
& & \times \, w(\vec{k}_{a\perp},\vec{k}_{b\perp}) \, X(x_a,\vec{k}_{a \perp}) \, Y(x_b,\vec{k}_{b \perp}) \,,
\label{e:convolution}
\end{eqnarray}
for the convolution of transverse parton momenta, with $w(\vec{k}_{a\perp}, \vec{k}_{b\perp})$ representing a generic weight factor. 
Note that in the case of the complex conjugate amplitude we have denoted the transverse gluon momenta by $\vec{p}_{a\perp}$, $\vec{p}_{b\perp}$.
Through the observable in Eq.~\eqref{e:obs_F14}, for the term which is exclusively given by F-type GTMDs, one ``selects" $F_{1,4}$ for the nucleon $N_a$.
The 3rd and 4th term on the r.h.s.~of this equation is given by G-type GTMDs only, while the last term arises from the interference of F-type and G-type GTMDs.
At present very little is known about the (relative) magnitude of the various gluon GTMDs in Eq.~\eqref{e:obs_F14}.
In order to arrive at a simplified expression for the observable we use the fact that in the forward limit (the real part of) $F_{1,1}$ reduces to the density of unpolarized gluons.
We therefore assume that $F_{1,1}$ is (much) larger than all the other GTMDs.
In turn, $G_{1,4}$ becomes the gluon helicity distribution in the forward limit, and we consider this function to be larger than the remaining GTMDs (except of course $F_{1,1}$), even though currently this distribution still has very large uncertainties~\cite{deFlorian:2014yva, Kovchegov:2017lsr}.
With these approximations one finds
\begin{eqnarray}
\lefteqn{\frac{1}{4} \big( \tau_{UU} + \tau_{LL} - \tau_{XX} - \tau_{YY} \big)}
\nonumber \\[0.1cm]
& \approx & \frac{1}{M^4} \, \big( \varepsilon_\perp^{ij} \Delta q_\perp^i \Delta_{a\perp}^j \big)^2 \,
C \Big[ \vec{\beta}_{\perp} \cdot \vec{k}_{a \perp} \, F_{1,4}(x_a, \vec{k}_{a\perp}) \, F_{1,1}(x_b, \vec{k}_{b\perp}) \Big] \,
C \Big[ \vec{\beta}_{\perp} \cdot \vec{p}_{a \perp} \, F_{1,4}^{\ast}(x_a, \vec{p}_{a\perp}) \, F_{1,1}^{\ast}(x_b, \vec{p}_{b\perp}) \Big]
\nonumber \\[0.1cm]
& + & C \Big[ G_{1,4}(x_a, \vec{k}_{a\perp}) \, G_{1,4}(x_b, \vec{k}_{b\perp}) \Big] \,
C \Big[ G_{1,4}^{\ast}(x_a, \vec{p}_{a\perp}) \, G_{1,4}^{\ast}(x_b, \vec{p}_{b\perp}) \Big] \,,
\label{e:obs_F14_approx}
\end{eqnarray}
where the vector $\vec{\beta}_\perp$ is given by 
\begin{equation}
\vec{\beta}_\perp = \frac{ \vec{\Delta}_{a\perp}^2 \, \Delta \vec{q}_\perp - (\vec{\Delta}_{a\perp} \cdot \Delta \vec{q}_\perp) \, \vec{\Delta}_{a\perp}}{\vec{\Delta}_{a\perp}^2 \, \Delta \vec{q}_\perp^{\; 2} - (\vec{\Delta}_{a\perp} \cdot \Delta \vec{q}_\perp)^2} 
= \frac{\vec{\Delta}_{a \perp} \times ( \Delta \vec{q}_\perp \times \vec{\Delta}_{a \perp})}{\big( \varepsilon_\perp^{ij} \Delta q_\perp^i \Delta_{a\perp}^j \big)^2} \,.
\label{e:beta}
\end{equation}  
It is not possible to disentangle in a model-independent manner the two terms on the r.h.s.~of Eq.~\eqref{e:obs_F14_approx}.
Since GTMDs also depend on the variable $\vec{k}_\perp \cdot \vec{\Delta}_\perp$, even the 2nd term on the r.h.s.~of~\eqref{e:obs_F14_approx} depends on the angle $\varphi$ between $\vec{\Delta}_{a \perp}$ and $\Delta \vec{q}_\perp$.
However, if this dependence is mild, one may be able to separate the two terms in Eq.~\eqref{e:obs_F14_approx} by measuring this observable as function of $\varphi$.

For the GTMD $G_{1,1}$, the observable corresponding to Eq.~\eqref{e:obs_F14} is 
\begin{eqnarray}
\lefteqn{\frac{1}{4} \big( \tau_{UU} + \tau_{LL} + \tau_{XX} + \tau_{YY} \big)}
\nonumber \\[0.1cm]
& = & \frac{\varepsilon_\perp^{ij} \Delta_{a \perp}^j}{M} \, \frac{\varepsilon_\perp^{kl} \Delta_{a \perp}^l}{M} \,
C \bigg[ \frac{k_{a \perp}^i}{M} G_{1,1}(x_a, \vec{k}_{a\perp}) \, G_{1,4}(x_b, \vec{k}_{b\perp}) \bigg] \,
C \bigg[ \frac{p_{a \perp}^k}{M} G_{1,1}^{\ast}(x_a, \vec{p}_{a\perp}) \, G_{1,4}^{\ast}(x_b, \vec{p}_{b\perp}) \bigg]
\nonumber \\[0.1cm]
& + & (1 - \xi_b^2)^2 \, \frac{\varepsilon_\perp^{ij} \Delta_{a \perp}^j}{M} \, \frac{\varepsilon_\perp^{kl} \Delta_{a \perp}^l}{M} \,
C \bigg[ \frac{k_{a \perp}^i k_{b \perp}^m}{M^2} G_{1,1}(x_a, \vec{k}_{a\perp}) \, G_{1,2}(x_b, \vec{k}_{b\perp}) \bigg] \,
C \bigg[ \frac{p_{a \perp}^k p_{b \perp}^m}{M^2} G_{1,1}^{\ast}(x_a, \vec{p}_{a\perp}) \, G_{1,2}^{\ast}(x_b, \vec{p}_{b\perp}) \bigg]
\nonumber \\[0.1cm]
& + & C \Big[ F_{1,1}(x_a, \vec{k}_{a\perp}) \, F_{1,1}(x_b, \vec{k}_{b\perp}) \Big] \,
C \Big[ F_{1,1}^{\ast}(x_a, \vec{p}_{a\perp}) \, F_{1,1}^{\ast}(x_b, \vec{p}_{b\perp}) \Big]
\nonumber \\[0.1cm]
& + & (1 - \xi_b^2)^2 \, C \bigg[ \frac{k_{b \perp}^m}{M} F_{1,1}(x_a, \vec{k}_{a\perp}) \, F_{1,2}(x_b, \vec{k}_{b\perp}) \bigg] \,
C \bigg[ \frac{p_{b \perp}^m}{M} F_{1,1}^{\ast}(x_a, \vec{p}_{a\perp}) \, F_{1,2}^{\ast}(x_b, \vec{p}_{b\perp}) \bigg]
\nonumber \\[0.1cm]
& - & \, 2 (1 - \xi_b^2)^2 \, \frac{\varepsilon_\perp^{ij} \Delta_{a \perp}^j}{M} \, \varepsilon_{\perp}^{mn} \, {\rm Re} \,
\bigg\{ C \bigg[ \frac{k_{a \perp}^i k_{b \perp}^m}{M^2} G_{1,1}(x_a, \vec{k}_{a\perp}) \, G_{1,2}(x_b, \vec{k}_{b\perp}) \bigg] \,
C \bigg[ \frac{p_{b \perp}^n}{M} F_{1,1}^{\ast}(x_a, \vec{p}_{a\perp}) \, F_{1,2}^{\ast}(x_b, \vec{p}_{b\perp}) \bigg] \bigg\} \,. \qquad 
\label{e:obs_G11}
\end{eqnarray}
While we are interested in the 1st term on the r.h.s.~of Eq.~\eqref{e:obs_G11}, based on the above discussion about the magnitude of gluon GTMDs we expect that the 3rd term clearly dominates this observable.
It therefore seems impossible to study $G_{1,1}$ through this observable.
This situation changes only if one polarizes the nucleon $N_b$ as well.

Like for the double Drell-Yan process presented in~\cite{Bhattacharya:2017bvs} we also consider observables which depend on the interference of $F_{1,4}$ and $G_{1,1}$ with other GTMDs for the nucleon $N_a$ that are expected to be large.
This situation occurs for 
\begin{eqnarray}
\lefteqn{\frac{1}{2} \big( \tau_{UL} + \tau_{LU} \big)}
\nonumber \\[0.1cm]
& = & 2 \, {\rm Im} \, \bigg\{ - \frac{\varepsilon_\perp^{ij} \Delta_{a \perp}^j}{M} \,
C \bigg[ \frac{k_{a \perp}^i}{M} F_{1,4}(x_a, \vec{k}_{a\perp}) \, F_{1,1}(x_b, \vec{k}_{b\perp}) \bigg] \,
C \Big[ F_{1,1}^{\ast}(x_a, \vec{p}_{a\perp}) \, F_{1,1}^{\ast}(x_b, \vec{p}_{b\perp}) \Big]
\nonumber \\[0.1cm]
& - & (1 - \xi_b^2)^2 \, \frac{\varepsilon_\perp^{ij} \Delta_{a \perp}^j}{M} \,
C \bigg[ \frac{k_{a \perp}^i k_{b \perp}^m}{M^2} F_{1,4}(x_a, \vec{k}_{a\perp}) \, F_{1,2}(x_b, \vec{k}_{b\perp}) \bigg] \,
C \bigg[ \frac{p_{b \perp}^m}{M} F_{1,1}^{\ast}(x_a, \vec{p}_{a\perp}) \, F_{1,2}^{\ast}(x_b, \vec{p}_{b\perp}) \bigg]
\nonumber \\[0.1cm]
& + & \frac{\varepsilon_\perp^{ij} \Delta_{a \perp}^j}{M} \,
C \bigg[ \frac{k_{a \perp}^i}{M} G_{1,1}(x_a, \vec{k}_{a\perp}) \, G_{1,4}(x_b, \vec{k}_{b\perp}) \bigg] \,
C \Big[ G_{1,4}^{\ast}(x_a, \vec{p}_{a\perp}) \, G_{1,4}^{\ast}(x_b, \vec{p}_{b\perp}) \Big]
\nonumber \\[0.1cm]
& + & (1 - \xi_b^2)^2 \, \frac{\varepsilon_\perp^{ij} \Delta_{a \perp}^j}{M} \,
C \bigg[ \frac{k_{a \perp}^i k_{b \perp}^m}{M^2} G_{1,1}(x_a, \vec{k}_{a\perp}) \, G_{1,2}(x_b, \vec{k}_{b\perp}) \bigg] \,
C \bigg[ \frac{p_{b \perp}^m}{M} G_{1,4}^{\ast}(x_a, \vec{p}_{a\perp}) \, G_{1,2}^{\ast}(x_b, \vec{p}_{b\perp}) \bigg]
\nonumber \\[0.1cm]
& + & (1 - \xi_b^2)^2 \varepsilon_\perp^{mn} \,
C \bigg[ \frac{k_{b \perp}^m}{M} F_{1,1}(x_a, \vec{k}_{a\perp}) \, F_{1,2}(x_b, \vec{k}_{b\perp}) \bigg] \,
C \bigg[ \frac{p_{b \perp}^n}{M} G_{1,4}^{\ast}(x_a, \vec{p}_{a\perp}) \, G_{1,2}^{\ast}(x_b, \vec{p}_{b\perp}) \bigg]
\nonumber \\[0.1cm]
& - & (1 - \xi_b^2)^2 \frac{\varepsilon_\perp^{ij} \Delta_{a \perp}^j}{M} \frac{\varepsilon_\perp^{kl} \Delta_{a \perp}^l}{M} \varepsilon_\perp^{mn} \,
C \bigg[ \frac{k_{a \perp}^i k_{b \perp}^m}{M^2} F_{1,4}(x_a, \vec{k}_{a\perp}) \, F_{1,2}(x_b, \vec{k}_{b\perp}) \bigg] \,
C \bigg[ \frac{p_{a \perp}^k p_{b \perp}^n}{M^2} G_{1,1}^{\ast}(x_a, \vec{p}_{a\perp}) \, G_{1,2}^{\ast}(x_b, \vec{p}_{b\perp}) \bigg]
\bigg\} \,, \qquad
\label{e:obs_interf}
\end{eqnarray}
where in the 1st and 2nd term on the r.h.s.~$F_{1,4}$ interferes with $F_{1,1}$ of the nucleon $N_a$, while in the 3rd and 4th term $G_{1,1}$ interferes with $G_{1,4}$.
Applying the aforementioned hierarchy of GTMDs one arrives at 
\begin{eqnarray}
\lefteqn{\frac{1}{2} \big( \tau_{UL} + \tau_{LU} \big)}
\nonumber \\[0.1cm]
& \approx & 2 \, {\rm Im} \, \bigg\{ - \frac{1}{M^2} \, \big( \varepsilon_\perp^{ij} \Delta q_\perp^i \Delta_{a\perp}^j \big) \,
C \Big[ \vec{\beta}_\perp \cdot \vec{k}_{a \perp} \, F_{1,4}(x_a, \vec{k}_{a\perp}) \, F_{1,1}(x_b, \vec{k}_{b\perp}) \Big] \,
C \Big[ F_{1,1}^{\ast}(x_a, \vec{p}_{a\perp}) \, F_{1,1}^{\ast}(x_b, \vec{p}_{b\perp}) \Big]
\nonumber \\[0.1cm]
& + & \frac{1}{M^2} \, \big( \varepsilon_\perp^{ij} \Delta q_\perp^i \Delta_{a\perp}^j \big) \,
C \Big[ \vec{\beta}_\perp \cdot \vec{k}_{a \perp} \, G_{1,1}(x_a, \vec{k}_{a\perp}) \, G_{1,4}(x_b, \vec{k}_{b\perp}) \Big] \,
C \Big[ G_{1,4}^{\ast}(x_a, \vec{p}_{a\perp}) \, G_{1,4}^{\ast}(x_b, \vec{p}_{b\perp}) \Big]
\bigg\} 
\nonumber \\[0.1cm]
& \approx & 2 \, {\rm Im} \, \bigg\{ - \frac{1}{M^2} \, \big( \varepsilon_\perp^{ij} \Delta q_\perp^i \Delta_{a\perp}^j \big) \,
C \Big[ \vec{\beta}_\perp \cdot \vec{k}_{a \perp} \, F_{1,4}(x_a, \vec{k}_{a\perp}) \, F_{1,1}(x_b, \vec{k}_{b\perp}) \Big] \,
C \Big[ F_{1,1}^{\ast}(x_a, \vec{p}_{a\perp}) \, F_{1,1}^{\ast}(x_b, \vec{p}_{b\perp}) \Big] \bigg\} \,.
\label{e:obs_interf_approx}
\end{eqnarray}
In the 2nd member of Eq.~\eqref{e:obs_interf_approx} we have kept two interference terms. 
However, since $F_{1,4}$ is accompanied by three powers of $F_{1,1}$ we expect this term to (clearly) dominate over the one containing $G_{1,1}$.
The external dependence on $\varphi$ --- determined through the prefactor $\varepsilon_\perp^{ij} \Delta q_\perp^i \Delta_{a\perp}^j$ and the vector $\vec{\beta}$ --- is identical for the two interference terms.
Hence, measuring the observable in~\eqref{e:obs_interf_approx} as function of $\varphi$ most likely will not help to disentangle the two terms.
We therefore conclude that this observable would allow one to study $F_{1,4}$ but not $G_{1,1}$.
Note however that in Eq.~\eqref{e:obs_interf_approx} appears the imaginary part of products of GTMDs, while at present the main interest is in ${\rm Re} \, F_{1,4}$ (and ${\rm Re} \, G_{1,1}$).
Instead of the observable in~\eqref{e:obs_interf_approx} one can therefore consider $\frac{1}{2} \big( \tau_{XY} - \tau_{YX} \big)$.
The result for this polarization observable is identical to the r.h.s.~of~\eqref{e:obs_interf_approx}, but with an overall minus sign and $ {\rm Re} \, ( \ldots )$ instead of ${\rm Im} \, ( \ldots )$.
In that case ${\rm Re} \, F_{1,4}$ is multiplied by the large ${\rm Re} \, F_{1,1}$ --- see the paragraph after Eq.~\eqref{e:convolution}.

\section{Summary} 
\label{s:concl}
We have shown that GTMDs of gluons can be measured through exclusive double production of pseudoscalar quarkonia ($\eta_{c}$ or $\eta_{b}$) in nucleon-nucleon collisions.
To this end, we have performed a LO analysis in perturbative QCD.
In that case gluon GTMDs in the ERBL region enter.
We have largely concentrated on two GTMDs ($F_{1,4}$ and $G_{1,1}$) which have attracted the most attention so far due to their intimate relation to the spin structure of the nucleon.
We have considered observables for the polarization of one initial-state nucleon plus recoil polarimetry of the corresponding final-state nucleon.
Our results indicate that via such observables $F_{1,4}$ can in principle be studied, while most likely there is not sufficient sensitivity to $G_{1,1}$.

We note that other polarization observables would allow one to address additional leading-twist gluon GTMDs through the same process.
In particular, polarizing also the second nucleon --- called $N_b$ above --- opens up further possibilities, including access to $G_{1,1}$. 
\\
\begin{acknowledgments}
One of us (A.M.) would like to thank F.~Petriello for a discussion about the exclusive production of quarkonium pairs at the LHC.
This work has been supported by the National Science Foundation under Contract No.~PHY-1516088 (S.B., A.M., J.T.), the National Science Foundation of China under Grant No.~11675093, and by the Thousand Talents Plan for Young Professionals (V.K.O., J.Z.).
\end{acknowledgments}

\end{document}